\documentclass[1p]{elsarticle}

\usepackage{subfig}
\usepackage{caption}
\usepackage{hyperref}

\usepackage{amsmath}
\usepackage{booktabs}
\usepackage{graphicx}
\usepackage{multirow}
\usepackage{courier}
\usepackage{multirow}

\usepackage[colorinlistoftodos,prependcaption,textsize=small]{todonotes}
\usepackage{blindtext}
\presetkeys%
    {todonotes}%
    {inline,backgroundcolor=yellow}{}
\graphicspath{{.}{images/}}
\DeclareGraphicsExtensions{.pdf}

\journal{Simulation Modelling Practice and Theory}

\begin{document}

\begin{frontmatter}

\title{Multi-level Simulation of\\ Internet of Things on Smart Territories \footnotemark[0] \footnotemark[1]}

\footnotetext[0]{Some parts of the research work described in this paper previously appeared in~\cite{gda-hpcs-16,smartshires,smartshires_abps}. This paper is an extensively revised and extended version of the previous work in which more than 30\% is new material. Please note that, some parts of the performance evaluation appeared in the extended version of~\cite{smartshires} that is available in~\url{https://arxiv.org/abs/1604.07076} but not included in the official conference proceedings.}

\footnotetext[1]{The publisher version of this paper is available at \url{http://dx.doi.org/10.1016/j.simpat.2016.10.008}.
\textbf{{\color{red}Please cite this paper as: ``Gabriele D'Angelo, Stefano Ferretti, Vittorio Ghini. Multi-level Simulation of Internet of Things on Smart Territories. Simulation Modelling Practice and Theory, Elsevier, vol. 73 (April 2017)''.}}}

\author{Gabriele D'Angelo\corref{cor1}}
\ead{g.dangelo@unibo.it}

\author{Stefano Ferretti}
\ead{s.ferretti@unibo.it}

\author{Vittorio Ghini}
\ead{vittorio.ghini@unibo.it}

\address{Department of Computer Science and Engineering. University of Bologna, Italy.}

\cortext[cor1]{Corresponding Author. Address: Department of Computer Science and Engineering. University of Bologna. Mura Anteo Zamboni 7. I-40127, Bologna. Italy. Phone +39 051 2094511, Fax +39 051 2094510} 

\begin{abstract}
In this paper, a methodology is presented and employed for simulating the Internet of Things (IoT).
The requirement for scalability, due to the possibly huge amount of involved sensors and devices, and the heterogeneous scenarios that might occur, impose resorting to sophisticated modeling and simulation techniques.
In particular, multi-level simulation is regarded as a main framework that allows simulating large-scale IoT environments while keeping high levels of detail, when it is needed. 
We consider a use case based on the deployment of smart services in decentralized territories. A two level simulator is employed, which is based on a coarse agent-based, adaptive parallel and distributed simulation approach to model the general life of simulated entities. However, when needed a finer grained simulator (based on OMNeT++) is triggered on a restricted portion of the simulated area, which allows considering all issues concerned with wireless communications. 
Based on this use case, it is confirmed that the ad-hoc wireless networking technologies do represent a principle tool to deploy smart services over decentralized countrysides.
Moreover, the performance evaluation confirms the viability of utilizing multi-level simulation for simulating large scale IoT environments.
\end{abstract}


\end{frontmatter}

\section{Introduction}
\label{sec:intro}

The Internet is growing at an incredible pace. 
A multitude of mobile users' terminals, sensors, RFID devices, and ``things'' in general has been (and it is going to be) designed to offer novel services in smart cities and territories~\cite{iot_survey,Atzori:2010,gda-hpcs-11}.
All these devices vary in terms of hardware and software characteristics. 
The variety of these devices and the services that form such Internet of Things (IoT) represents a very interesting opportunity with an important impact on social good initiatives around the world.

In this extremely heterogeneous scenario, it becomes essential being able to understand and to simulate the IoT. 
The complex networks obtained by the interaction of IoT devices are hard to design and to manage. In real deployment scenarios, many possible configurations of IoT networks are possible. Devices’ connectivity is influenced by their geographical location, distribution, communication and network capabilities~\cite{gda-hpcs-16}.
IoT simulation is necessary for both quantitative and qualitative aspects. To name a few issues: capacity planning, what-if simulation and analysis, proactive management and support for many specific security-related evaluations. 
The scale of the IoT is the main problem in the usage of existing simulation tools. Traditional approaches (that are single CPU-based) are often unable to scale to the number of nodes (and level of detail) required by the IoT. 

In this paper we propose a technique to simulate IoT using a multi-level simulation approach~\cite{Ghosh:1986:CDM:319541.319559}. This approach is able to provide scalability and real-time execution of massively populated IoT environments without forcing the simulationist to over-simply the models.
In essence, in a multi-level simulation multiple simulation models are glued together, each one with a specific task and working at a different level of detail~\cite{magne2000towards}.
The idea is to exploit a ``high level'' simulator that works at a coarse grained level of detail. This coordinates the execution of a set of domain specific ``lower level'' simulators that are used only when a fine grained level of detail is necessary. 
Such lower level simulators can in turn trigger further simulations that act at an even finer level of detail, and so on.
Based on the specific simulation, care should be put in the interoperability among the simulators and the design of the inter-model interactions (e.g.~synchronization and state exchanges at runtime between model components).

We demonstrate the validity of the proposed approach by focusing on the simulation of ``smart territories''~\cite{smartshires,smartshires_abps}.
This is a novel view of devising smart services over urban and decentralized environments. Indeed, in these last months/years focus has been given on the development of smart cities services, i.e.~a set of strategies aiming at improving and optimizing services offered to citizens living in metropolitan areas.
As a matter of fact, the possibility to offering services for territorial districts with low population density is an almost ignored problem. 
% We are not referring here to the well-known ``digital divide'', a gap that has been widely recognized and that, in certain areas of more civilized countries (e.g., Europe), is being solved, more or less rapidly. Several (non-metropolitan) territories, composed of small towns, can offer modern (yet, by now, traditional) networking infrastructures. However, surrounding areas are commonly not covered by further communication services, besides network connectivity. Indeed, to follow the smart cities trend, connectivity is a necessary but not sufficient feature.
Thus, the idea here is to create a geographical space able to manage resources (natural, human, equipment, buildings and infrastructure) in a way that is sustainable and not harmful to the environment. 
This goal to bridge the urban-rural divide needs means to leverage novel wireless and mobile technologies, together with smart computing services. In fact, smart services would make good use of a deployment of cheap sensors in these areas, together with ad-hoc configurations of mobile devices, without the need of costly (communication) infrastructures.
Examples of technologies that might compose the substrate for developing a ``smart shire'' platform are: multihoming mobile services, mobile ad hoc networks, opportunistic networks, peer-to-peer, cloud and fog computing systems.
A smart middleware can be built that leverages these services; such a middleware would simplify the development of new services and the
integration of legacy technologies into new ones~\cite{Atzori:2010}. On top of the middleware, smart services will be deployed, whose application domains include information dissemination, tourist services in areas where a communication infrastructure is not present, info-mobility, weather services, pollution in rural areas (e.g., dumps), sustainable (sub)urban environment, healthy services for ageing population, continuous care, emergency response, smart renewable energy.

We show that the design and configuration of smart services in (decentralized) territories impose the simulation of wide area networks; however, in certain cases a highly detailed simulation is required. This need for scalability and high level of detail can be reached only through properly configured multi-level simulation techniques.
An advantage of this approach is that the detailed (and thus, more costly) simulation can be performed only when needed, in a limited simulated area, only for the needed time interval of the simulation.

As a use case, we describe the implementation of a multi-level approach to simulate a smart market scenario.
The market is composed of several producers that publish availability of offered products. Interested clients can subscribe to these advertisements and go to the market to buy these (and other products). Let imagine that the market is particularly crowded and the client might desire to have further information on presence of possible alternative products and position of the producers. 
Thus, producers might provide information on the fly, thanks to smart proximity-based services that guide customers through market.
Different wireless technologies can be exploited to deploy such a kind of proximity-based applications.

In this scenario, while the first part of the interaction can be modeled through a classic agent-based simulation, the proximity-based interactions require finer grained simulation details, involving wireless communication protocols. In this case, we show that the use of multi-level simulation provides means to define and manage the whole picture in a proper way.

% An important methodology to preliminarily assess the viability of large scale solutions in wide areas, such as countrysides, is simulation. In this case, simulation must be carefully handled in order to create reasonably accurate models that can scale in terms of modeled entities and granularity of events. Indeed, even a small size smart shire will be composed by thousands of (possibly) interconnected sensors and devices. Thus, scalability is the main requirement to consider. Scalability would allow considering smart shires that are connected to smart cities, thus creating a whole smart territory. To this aim, probably the best approach accounts to the use of discrete-event simulation combined with an agent-based model. 
% 
% With this aim, we have developed a Smart Shire Simulator ($S^3$), an agent based model built on top of GAIA/ART\`IS, that is a simulation middleware enabling the seamless sequential/parallel/distributed execution of large scale simulation runs. Using $S^3$, we study the effectiveness of a ``priority-based broadcast'' dissemination scheme employed over ad-hoc networks, where communication among nodes is made possible through direct transmissions among near devices. Results show that dissemination schemes (when coupled with caching schemes) can represent an effective communication substrate to be used in a software middleware promoting the creation of applications for smart shire scenarios.

We provide results from an experimental assessment demonstrating that multi-level simulation offers a better scalability with respect to a classic fine grained simulation, while offering the same level of detail, when needed.

The remainder of this paper is organized as follows. 
Section~\ref{sec:smart-territories} describes the main issues concerned with the development of smart shires and smart territories. 
This preliminary discussion allows introducing the main issues that need to be addressed when employing simulation in this described scenario.
Section~\ref{sec:multilevel} introduces multi-level simulation. 
Section~\ref{sec:simulator} provides a discussion on the approach we propose to simulate large scale IoT environments on smart territories.
In Section~\ref{sec:perf} a preliminary performance evaluation of the implemented simulators is shown.
Section~\ref{sec:related} discusses on related work.
Finally, Section~\ref{sec:conclusions} provides some concluding remarks.

\section{Smart Territories}
\label{sec:smart-territories}

Pervasive computing technologies represent a novel means to offer services that may have a big impact and add several benefits to citizens in their daily activities. 
It is a fact that the current trend is to devise and deploy services in metropolitan and crowded areas, based on the idea that the higher the amount of possible users, the higher the amount of people that would benefit from these services (and the higher the amount of clients).

However, recent proposals claim a need to put focus on the development of cheap and sustainable services for non-metropolitan areas and countrysides in general~\cite{smartshires,smartshires_abps}.
The goal is to promote and take advantage of the potential of decentralized areas via deployment of smart services, which would be probably different to those that can be deployed in metropolitan areas. Rather, self-configuring opportunistic solutions should be devised, possibly not strictly dependent to the presence of a classic urban services and networking infrastructure. 

\subsection{The need for smart services in smart shires}

Making a territory smart involves leading innovative solutions on countrysides; we call this novel view of equipping smart services in countrysides ``smart shires''.
The domain of services that may provide some benefit to these areas is quite broad. 
They range from services to citizens to services for municipalities.
Examples of services for citizens are improved Internet access, digital municipalities, apps promoting citizen participation, geo-referenced information for a multitude of user applications (e.g., for tourists). A main use case for smart territories in general is that of proximity-based applications, where devices detect their proximity and subsequently trigger different services.

Services might allow federating neighbor municipalities/towns, private or public organizations so as to form a critical mass of both offered services and potential users. In this scenario, wireless networks, with specific focus on ad-hoc, 5G Device-to-Device (D2D) communications and multihoming, have great significance for the organization of a smart territory.

It makes sense to consider the scenario of various heterogeneous devices interconnected one to each other and to exploit these interconnections to create novel services~\cite{Petrolo:2014}. Data sensed by the sensors' devices are disseminated and collected by an information processing system, managed as open data within the middleware, to be used by applications. Such a middleware should enable a context-aware data distribution, i.e.~it should be able to distribute context data to interested entities~\cite{Bellavista:2012}.

\subsection{Enabling technologies for smart territories/shires}

Several enabling technologies can be put to good use to build effective services in smart shires and smart territories in general.
The main constraint to consider is the need for cheap and sustainable solutions.
IoT, mobile and pervasive computing, crowd-sourcing and crowd-sensing are thus main pieces of the puzzle.

Sensors are relatively cheap in terms of costs. Thus, their deployment in a countryside is feasible. Problems may arise to interconnect these sensors to form a sensor network and to make them communicate with the intelligent services placed in the Internet. This requires the use of smart services employing D2D and multi-hop/multi-path communications~\cite{Wirtz:2014}.

\subsubsection{Communication strategies}

The multitude of data generated by sensors, users, public and private organizations, to be made available for a plethora of possible services requires a careful management and dissemination of data. The standard client/server approach might be useful in certain contexts, but other dissemination strategies have to be made available.
Smart territories should be able to exploit pervasive solutions, where IoT, wireless sensor networks and ubiquitous computing are merged in a single platform. 

\begin{figure}[t]
\centering
 \includegraphics[width=.7\linewidth]{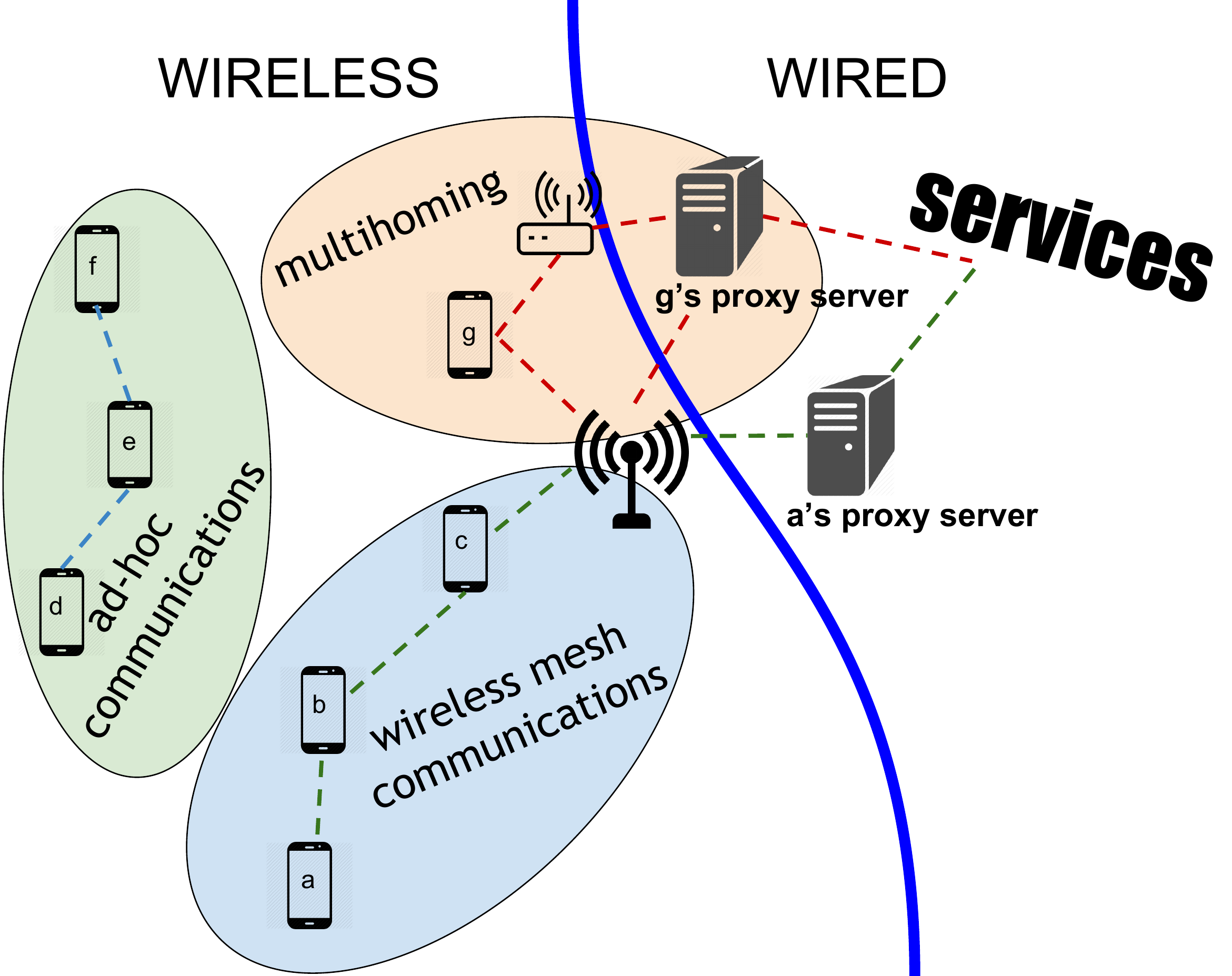}
 % archi_logic.eps: 0x0 pixel, 300dpi, 0.00x0.00 cm, bb=14 14 610 856
\caption{Multipoint and multihoming communications.}
\label{fig:mhop}
\end{figure}

Figure~\ref{fig:mhop} summarizes the different types of communications that may arise in the considered smart shire scenarios. In the figure, nodes $d, e$ and $f$ perform an ad-hoc communication, without the intervention of a network infrastructure. Node $a$ is able to connect to Internet services via a wireless mesh network, that exploits a multi-hop communication to let messages reaching a network infrastructure~\cite{Alotaibi2012940}. Finally, $h$ connects to Internet services thanks to the use of multi-homing, that allows utilizing its multiple network interface cards concurrently, in a seamless way~\cite{Ferretti2016390}. 
This means ensuring that if a mobile node changes its point of attachment to the Internet, while in movement, no communication interruptions are perceived at the application level, and if such interruptions occur, they do not significantly degrade the Quality of Service delivered at the application level~\cite{GhiniJSS}.

\subsubsection{Dissemination strategies}

At a different, higher level of abstraction, general approaches such as publish/subscribe and gossip-style dissemination schemes can be exploited~\cite{smartshires,Ferretti20131631,carretero:hal-00764234}.
Depending on the application scenario a message might have to be sent from a device towards a certain physical area of the territory (e.g., the device is looking for a sensor to understand weather conditions) or towards a specific destination (as in classic MANET/VANET scenarios), or the message ``is looking for an access point'', trying to reach an Internet host or service. In other cases, applications might require a broadcast of a message; this is the case for alert messages, critical situations, or more simply general information or advertisements. To this aim, an efficient broadcast scheme might be employed that spread messages across devices, trying to avoid transmissions' collisions. The broadcast might contain aggregate contents (marshalling), in order to optimize transmissions.
This provides a useful scheme to be employed at higher levels and disseminate requests and contents~\cite{smartshires_abps}.

As concerns publish-subscribe strategies, there are several alternatives.
For example, when dealing with publish-subscribe for pervasive computing and IoT, it is worth citing MQ Telemetry Transport (MQTT), a lightweight broker-based publish/subscribe messaging protocol~\cite{mqtt}. 
Its simplicity makes it an ideal messaging protocol for the IoT and mobile communications. 
Another option is the Advanced Message Queuing Protocol (AMQP), an application layer protocol for the IoT
focusing on message-oriented environments~\cite{amqp}. 
Data Distribution Service (DDS) is a publish-subscribe protocol, developed by Object Management Group (OMG), that relies on multicasting, through a broker-less architecture~\cite{dds}.

\subsection{A use case with smart marketplaces}

Let consider an application scenario concerned with the ``km 0'' phenomenon. This is the abbreviation for ``zero kilometers'', that signifies local, low impact primary food ingredients. It refers to a shopping style that gained a lot of interests in Italian and European foodie circles to improve the quality of products and promote sustainable cooking.
It gives priority to the use of local and seasonal foods, avoiding genetically modified organisms. In spite of the growing interest in local products, there are relatively few places where one can buy these products directly from the producer. Thus, customers have to look for specialized weekend farmers' markets or for farm direct purchases. Customers might be single users, ethical purchasing groups, restaurant owners. And quite often, this products' research reveals to be not a simple task for customers. 

Thus, let imagine a service that allows consumers subscribing to the availability of a certain product. Upon availability, a producer (e.g.~the farmer) can notify his subscribers of product availability plus other related information such as, for instance, his presence in next, nearby markets or other possible purchasing opportunities.
In view of such details, the consumer can plan visiting the market (and select the products directly), book some specific items, quantities and so on. 
The plethora of publish-subscribe systems available can easily do this job. 
However, more sophisticated services are possible. 
Several (apparently similar) producers can be present at the market, the customer might do not know the location, he might have some physical disabilities and thus he might need to be guided to the exact location of the producer, that is dynamically determined (hence, without the possibility of knowing the position in advance).
Then, once there, he might be possibly interested in finding other products.
Moreover, once there, he might want checking out for other interesting stands.

Thus, it would be an added value if producers might provide information on the fly, thanks to smart proximity-based services that could guide customers through the market.
Such services can be deployed by adapting the interactions and the communications among end-nodes based on the locally available communication technologies. For instance, in presence of a wireless infrastructure, all the communications can pass through the access point and the Internet. Otherwise, some ad-hoc solution should be dynamically built, with producers that exploit their smart devices (e.g.~smartphones) to build multihop wireless communication and information dissemination strategies~\cite{smartshires_abps}.
Moreover, in case of intermittent connections, seamless (multihoming based) communication strategies should be employed~\cite{Ferretti2016390}.
Being partly composed of advertisements, general information on the market, published messages looking for their subscribers, dissemination over such dynamic, opportunistic ad-hoc overlay might be based on epidemic dissemination protocols, used in conjunction with application filtering techniques~\cite{Ferretti2013481,Wirtz:2014}.

The efficient simulation of such a wide scenario in a smart territory is not an easy task, since it involves several activities and different domains. 
This is a perfect example of a simulation scenario requiring different levels of granularity. Thus, it is necessary to employ multi-level simulation. 
This methodology is described in the next section.

\section{Multi-level Simulation Architecture}
\label{sec:multilevel}

This section describes multi-level simulation, an approach for large scale simulation setups~\cite{gda-hpcs-16,Ghosh:1986:CDM:319541.319559}. The rationale is to take multiple simulation models, glued together into a hierarchical structure, each one with a specific task and working at a different level of detail~\cite{magne2000towards}.
The simulation starts with the ``high level'' simulator that works at a coarse grained level of detail.
Such a simulator coordinates the execution of a set of domain specific ``middle'' or ``low'' level simulators that are used only when a fine grained level of detail is necessary. The switch between coarse and fine grained models can be automatic or triggered by the simulation modeler. 
For example, imagine to have an agent-based simulator that works at a coarse level of detail (this is actually the solution employed in this work, and described in Section~\ref{sec:simulator}).
When it is required to simulate an area that, for instance, is populated by a large number of wireless devices, then a detailed simulation model could 
be employed and executed by typical simulators that have been designed for this purpose e.g.~OMNeT++~\cite{omnet}, ns-3~\cite{ns3}, SUMO~\cite{sumo}.

\begin{figure}[ht]
\centering
\includegraphics[width=\linewidth]{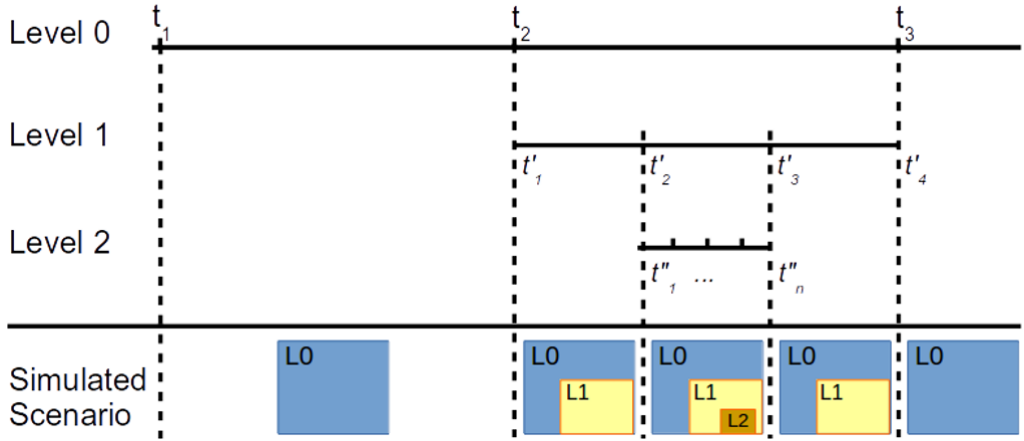}
\caption{Multi-level simulation.}
\label{fig:multilevel}
\end{figure}
%%%%%%%%%%%%%

Figure~\ref{fig:multilevel} provides a overview of this approach~\cite{gda-hpcs-16}. At the simulation bootstrap the whole scenario is executed at Level 0 (that is, with minimal details). Hence, the high level simulator manages the evolution of all the model components and their interactions following a time-stepped synchronization approach.~\cite{gda-simpat-2014}.
At timestep $t_2$, it is found that a part of the simulated scenario (for example a specific zone in the simulated area or a specific group of modeled nodes) has to be simulated with more details. This means that, in the figure, a part of the simulated area is still modeled at Level 0 while a specific zone is now managed by the Level 1 model. If necessary, in the following of the simulation, a specific area can be further detailed using a Level 2 model (and simulator). To simplify this discussion, if we consider only two levels, then all the model components managed by the Level 0 simulator are evolved using $t$-sized (coarse grained) timesteps and all the others use $t'$-sized (fine grained) timesteps. Timestep $t_2$ (that is the same of $t'_1$ for Level 1) is the moment in which a part of the model components is transferred from the coarse grained simulator to the finer one. In the following, the components at Level 0 will jump from $t_2$ to $t_3$ while the components simulated at Level 1 will be updated at $t'_2$, $t'_3$ and $t'_4$ (that is the same of $t_3$ for Level 0). Then, since there is no more need for such a level of detail, all the components simulated at Level 1 are transferred again to the Level 0 simulator. Following the constraints imposed by the time-stepped synchronization algorithm, all the interactions among Level 0 simulated components can happen at every coarse grained timestep while the interactions at Level 1 can happen at every fine grained timestep. Finally, the interaction between components managed at different levels can happen only at the coarse grained timesteps; that is, when there is a match between the timesteps at the different levels.

From the discussion above it should be clear that when designing a multi-level simulation approach, the main issues are the interoperability among the simulators and the design of the inter-model interactions (e.g.~synchronization and state exchanges at runtime between model components).
Moreover, while in Figure~\ref{fig:multilevel} both shown levels are following a time-stepped approach, it might be possible that different simulators, acting at different levels of detail, employ different simulation advancement approaches. For instance, in the implementation of the use case discussed in the next section, we will show that Level 0 (coarse grained) is a time-stepped simulator, while Level 1 (fine grained) is an event-driven OMNeT++ based simulator.
 
Multi-level simulation can be employed by the industry and the academia in a plethora of different contexts, ranging from scientific environments to manufacturing and industrial systems. Focusing on the IoT and smart cities, the novel trends in automation and manufacturing technologies are leading to what is called the fourth industrial revolution. The idea is that, currently, cyber-physical systems can communicate and cooperate with each other and with humans in real time, hence creating a modular and complex production chain that promotes interoperability, information transparency and decentralized decisions. The modular structure of this production chain takes into consideration several factors and issues, ranging from the use of small sensors to the use of big data in order to take management decisions. Multi-level simulation might be a perfect tool to understand how organizing and managing such a complex production chain. In this paper, we will show how multi-level simulation can be proficiently employed to simulate a smart territory, focusing on the smart-market use case scenario.

\section{Multi-level Simulator for Smart Territories}
\label{sec:simulator}

In this section, we describe the implementation of a multi-level approach to simulate the smart market scenario in smart shire environments, described in Section~\ref{sec:smart-territories}. 
The first, main requirement is scalability, in terms of modeled entities and granularity of events. Even a small size smart shire will be composed by thousands of interconnected devices. Many of them will be mobile and each with very specific behavior and technical characteristics. 
Another requirement is that the simulator should be able to run in (almost) real-time average size model instances. This enables proactive approach (e.g.~simulation in the loop) and to perform ``what-if analysis'' during the management of the deployed architecture. 

Considering the characteristics of the model to be simulated and the requirements described above, in our view the best way is to use a multi-level approach that combines a discrete-event simulation engine coupled with an agent-based smart shire model.

\begin{figure}[ht]
\centering
\includegraphics[width=\linewidth]{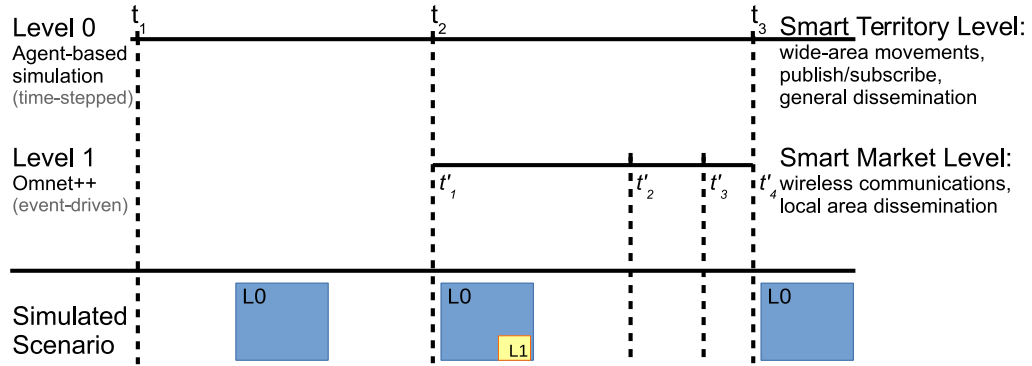}
\caption{Smart Territory/Market multi-level simulation.}
\label{fig:usecase-multilevel}
\end{figure}

The simulator exploits two levels of granularity, as shown in Figure~\ref{fig:usecase-multilevel}. The two levels of simulations are two sophisticated, but in some sense standard, simulators.
The coarse level (Level 0) simulates the whole smart territory, where different actors produce products, subscribe their interests, move towards different geographical areas. This has been implemented using an agent-based simulator equipped with parallel and distributed execution capabilities~\cite{gda-mospas-11}.

Then, the specific interactions within the smart market impose more simulation details to consider wireless communication issues, fine-grained interactions and movements. 
Thus, an instance of OMNeT++ simulation has been implemented (i.e.~Level 1 in Figure~\ref{fig:usecase-multilevel}). In this case, each simulation step of the coarse grained simulation layer (e.g., $t_3, t_4$ in Level 0, Figure~\ref{fig:usecase-multilevel}) is decomposed into a set of events at the fine grained layer, organized as a pending event list (Level 1). Following this approach, Level 1 is able to notify Level 0 with its simulation advancements.

In this case, the main issue is to provide means to let the two simulators interact. In fact, Level 0 has to trigger Level 1 when needed, passing some arguments to it that would serve as configuration and initialization parameters (see Figure~\ref{fig:usecase-multilevel}). Then, Level 1 must run for a certain amount of timesteps and at the end of each Level 0 timestep, it has to notify Level 0 with its outputs. In turn, Level 0 must ask Level 1 to continue its simulation or end the simulator.
The operating principles of the two simulators and their interface is described with more details in the next subsections.

\subsection{Level 0: agent-based simulator}

Smart Shire Simulator ($S^3$) is a prototypal simulator based on the GAIA/ART\`IS simulation middleware~\cite{gda-simpat-2014,pads}. ART\`IS permits the seamless sequential/parallel/distributed execution of large scale simulation runs using different communication approaches (e.g.~shared memory, TCP/IP, MPI) and synchronization methods (e.g.~time-stepped, conservative, optimistic). The GAIA part of the software tool aims to ease the development of simulation models with high level application program interfaces. Furthermore, it implements communication and computational load-balancing strategies (based on the adaptive partitioning of the simulation model), for reducing the simulation execution time.

The current version of $S^3$ is preliminary and implements only a limited set of functionalities. The many elements composing the smart shire are represented as a set of interacting entities. Some entities are static (e.g.~sensors, traffic lights and road signs) while the others (e.g.~cars and smart-phones) follow specific mobility models. All the entities in the simulated model are equipped with a wireless device. The interaction among entities is based on a ``Priority-based Broadcast'' (PbB) strategy that implements a probabilistic broadcast approach~\cite{1200529}. In PbB, every messages that is generated by a node is broadcasted to all the nodes that are in proximity of the sender. The message contains a Time-To-Live (TTL) to limit its lifespan and the forwarding is based on two conditions. The first is a probabilistic evaluation (i.e.~probabilistic broadcast) while the second is based on the distance between sender and receiver. In fact, to limit the number of forwarded messages, there is a message forward only if the distance between the nodes is larger than a given threshold. Under the implementation viewpoint, this can be done using a positioning system (e.g.~GPS) if available. Otherwise, the network signal level associated to each received message is used. As will be discussed in Section~\ref{sec:perf}, this approach can generate a very large number of duplicated messages. For this reasons, the tuning of PdB parameters is very important and furthermore a message caching mechanism has been added.

\subsection{Level 1: OMNeT++ simulator}
\label{sec:omnet}

The finer grained simulator was implemented using OMNeT++ v.~4.4.1, with the INET framework v.~2.3.0. It simulates a grid of fixed nodes (during the tests, a $10\times10$ grid was used), representing the market sellers. Each seller is equipped with a WiFi enabled technology. In the simulated scenario, no WiFi infrastructure was present, hence nodes organize themselves as a MANET exploiting DYMOUM~\cite{www_dymo-um}, an implementation of the Dynamic MANET On-demand (DYMO) routing protocol~\cite{ietf-manet-dymo-26}.

In such a MANET a number $N$ of mobile nodes, representing pedestrian users, was introduced by the higher Level 0 simulator. These $N$ nodes are equipped with a mobile device with a WiFi networking technology. Pedestrian users move at walking speed. The user application running on the mobile client broadcasts messages looking for the identifier of the specific seller. The seller replies with his geographical position. All these messages are delivered through the mentioned MANET routing protocol. Based on the provided position, the mobile user moves towards his destination.

\subsection{Interface between the two simulators}

The interface between the Level 0 and Level 1 simulators is in charge of their interoperability and synchronization. That is, communicating the inputs and outputs between the two simulators, as well as of triggering the ``continue the simulation'' or ``end of simulation'' commands sent, at the end of each Level 0 timestep, from Level 0 to Level 1. To this aim, a message-passing approach is utilized, which has been realized through the use of a TCP connection in which the Level 0 simulator plays the role of server while the Level 1 simulator works as the client. Then, Level 0 waits for messages at a properly configured socket open to a certain port. At the end of each Level 0 timestep, Level 1 sends a set of messages which describe its status and waits for a response. Level 0 receives the data sent by Level 1 and decides what message has to be sent to Level 1 (i.e. continue or end the lower level simulation). The TCP connection is maintained until the Level 0 simulator decides that the lower level simulation must end.

In essence, this is a simple strategy that allows interactions between the simulators without requiring a complete re-engineering of the simulators. In this approach, the higher levels simulator must be able to freeze the simulation of certain parts of the simulated scenario, waiting for updates from other sources. Moreover, lower level simulators should be enabled to obtain input from outside, and notify results outside. However, no knowledge on the external simulators are needed. This is an example demonstrating that existing products can be employed to create more complex multi-level simulations.

Going into some technical detail, the Level 0 establishes a different TCP connection for each given Level 1 simulation instance. 
Before starting a Level 1 instance, the Level 0 opens a listening TCP port and passes that port number as a command-line parameter of the Level 1 simulator.  Analogously, the Level 0 dedicates a different directory for each given Level 1 simulation instance. The Level 0 writes the description of the scenario to be simulated in a set of files inside the directory associated to that given Level 1 instance.  These two mechanisms ensure the isolation between the Level 1 instances.

\subsection{Multi-level granularity and simulation approaches}
The multi-level approach described before can be combined with even finer grained simulation or emulation (and even analytic models). Moreover, many different simulation paradigms can be used for the implementation of both the coarser (i.e.~Level 0) and the finer grained simulator (i.e. Level 1). 
If the coarser grained level simulator (i.e. Level 0) uses a time-stepped approach, then there are only two requirements that must be fulfilled by the Level 1 simulator. Firstly, when the Level 0 simulator spawns a Level 1 instance this must be able to evolve its simulation state up to the end of the next Level 0 timestep without any external interaction (that is, in isolation from the Level 0 simulator). Secondly, at the end of the Level 0 timestep, the Level 1 simulator must be able to provide to Level 0 a correct simulation state. 
If even the Level 1 simulator uses a time-stepped synchronization, then these requirements can be easily satisfied by choosing the appropriate sizes for both the simulators timesteps. In mathematical terms, the Level 0 timestep size must be a multiple of the Level 1 timestep size. On the other hand, if the Level 0 or Level 1 simulators use a time advancing scheme that is not time-stepped then synchronization barriers (or time checkpoints) can be used to coordinate the simulators.

\section{Performance Evaluation}
\label{sec:perf}

In this section, firstly we evaluate the scalability of both Level 0 and Level 1 simulators. Secondly, we assess the performance of the multi-level simulator.

All the results reported in this section are averages of multiple independent runs. This performance evaluation has been performed on a DELL R620 with 2 CPUs and 128 GB of RAM. Each CPU is a Xeon E-2640v2, 2 GHz, 8 physical cores. Each CPU core supports Hyper-Threading and therefore the number of logical cores is 32. The computer is equipped with Ubuntu 14.04.3 LTS, GAIA/ART\`IS version 2.1.0, OMNeT++ v.~4.4.1 (with the INET framework v.~2.3.0). $S^3$ and the OMNeT++ model used for the multi-level simulator will be freely available as source code in the next release of GAIA/ART\`IS~\cite{pads} or upon request.

%%%%%%%%%%%%%%%%%%%%%%%%%%%%%%%%%%%%%%%%%%%%%%%%%%%
\subsection{Level 0: agent-based simulator}
%%%%%%%%%%%%%%%%%%%%%%%%%%%%%%%%%%%%%%%%%%%%%%%%%%%

The performance evaluation of $S^3$ is based on a bidimensional toroidal space (with no obstacles) that is populated by a given number of devices called Simulated Entities (SEs). A subset of the SEs follows a Random Waypoint (RWP)~\cite{rwp} mobility model while the others are static. The interaction among SEs is based on proximity and implements the multi-hop dissemination protocol previously described. In Table~\ref{table:model} the main parameters of this performance evaluation are reported. With respect to the simulation model described in~\cite{smartshires}, this new version of the simulator implements a caching mechanism used by each node to discard (some of) the duplicated messages generated by the PbB dissemination scheme. This caching system is based on a LRU (Least Recently Used) replacement algorithm.

Under the IoT simulation viewpoint, it is obvious that the model parameters to be chosen are strictly dependent on the specific scenario characterized by the geographical and architectural issues of each specific smart shire deployment. In our view, this confirms that a simulation based approach is needed to support the design of the architecture and for the appropriate tuning of runtime parameters.

\begin{table}[h]
\begin{center}
	\begin{tabular}{ | l | p{4cm} |}
	\hline
  \textbf{Model parameter} & \textbf{Description/Value} \\ \hline
	Number of $SEs$ & [1000, 32000] \\ \hline
  Mobility of $SEs$ & 50\% Random Waypoint (RWP)\newline 50\% static  \\ \hline
  Speed of RWP & Uniform in the range [1,14] spaceunits/timestep\\ \hline
  Sleep time of RWP & 0 (disabled)\\ \hline
  Interaction range & 250 spaceunits\\ \hline
  Density of $SEs$ & 1 node every 10000 $spaceunits^2$\\ \hline
  Forwarding range & $>200$ spaceunits\\ \hline  
  Simulated time & 900 timeunits \\ \hline
  Simulation granularity & 1 timestep = 1 timeunit \\ \hline
  Time-To-Live (TTL) & 4 hops \\ \hline
  Dissemination probability (gossip) & 0.6 \\ \hline
  Cache size (positions) &  0 (disabled) or 256 \\ \hline
  \end{tabular}
\end{center}
\caption{Simulation model parameters.}
\label{table:model}
\end{table}

\begin{figure}[h]
\centering
\includegraphics[width=8.5cm]{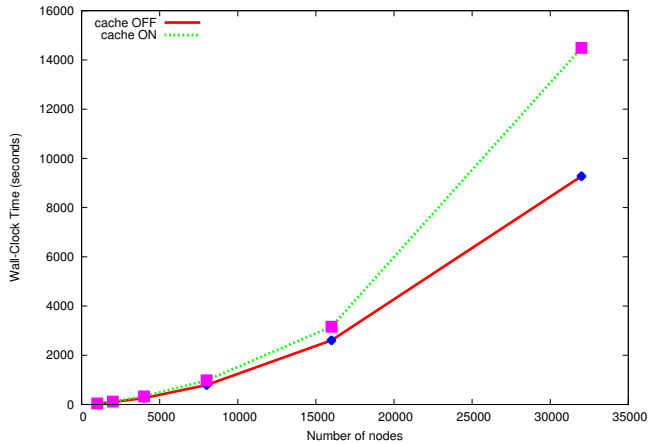}
\caption{Scalability evaluation: increasing number of SEs, sequential (\#CPU core=1) simulator, message caching OFF and ON.}
\label{fig_wct_seq}
\end{figure}

First of all, we assess the scalability of $S^3$ in a sequential setup (that is, 1 CPU core is used). Figure~\ref{fig_wct_seq} reports the Wall-Clock Time (WCT) to complete a single simulation with both message caching turned ON and OFF. The number of SEs has been set in the range [1000, 32000]. For each configuration, the size of the simulated area has been adjusted to maintain a fixed density of nodes. As expected, $S^3$ shows a good scalability but the performance are strongly affected by the number of SEs. 

The figure shows that the message caching mechanism introduces a significant overhead in terms of WCT and strongly affects the scalability of the simulator. We would have expected that the additional computational overhead introduced by caching would be balanced by the reduction in the overhead due to delivered duplicated messages, but this does not happen. 

\begin{figure}[h]
\centering
\includegraphics[width=8.5cm]{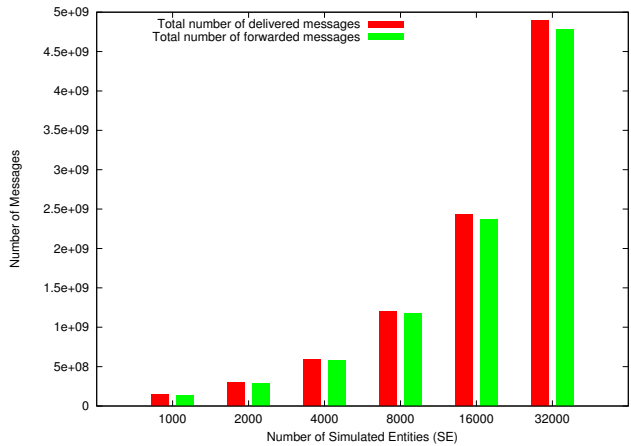}
\caption{Number of delivered messages in a single simulation run (cache OFF): increasing number of SEs, total delivered messages vs. forwarded messages}
\label{fig_messages}
\end{figure}

More in detail, Figure~\ref{fig_messages} shows the total number of messages (i.e.~interactions among devices) for each scenario. It results that even few SEs generate a very high number of messages and most of them are duplicated (forwarded) messages. In other words, most of the network traffic is overhead, since each new message is forwarded multiple times. The effect of this model behavior is a limitation in the simulator scalability and a clear design problem for the smart shire architecture. As a first effort to fix this problem, we have added to $S^3$ a message caching mechanism that operates in every node. This is an approach that is very common in presence of dissemination algorithms based on gossip~\cite{gda-jpdc-2017}.

\begin{figure}[h]
\centering
\includegraphics[width=8.5cm]{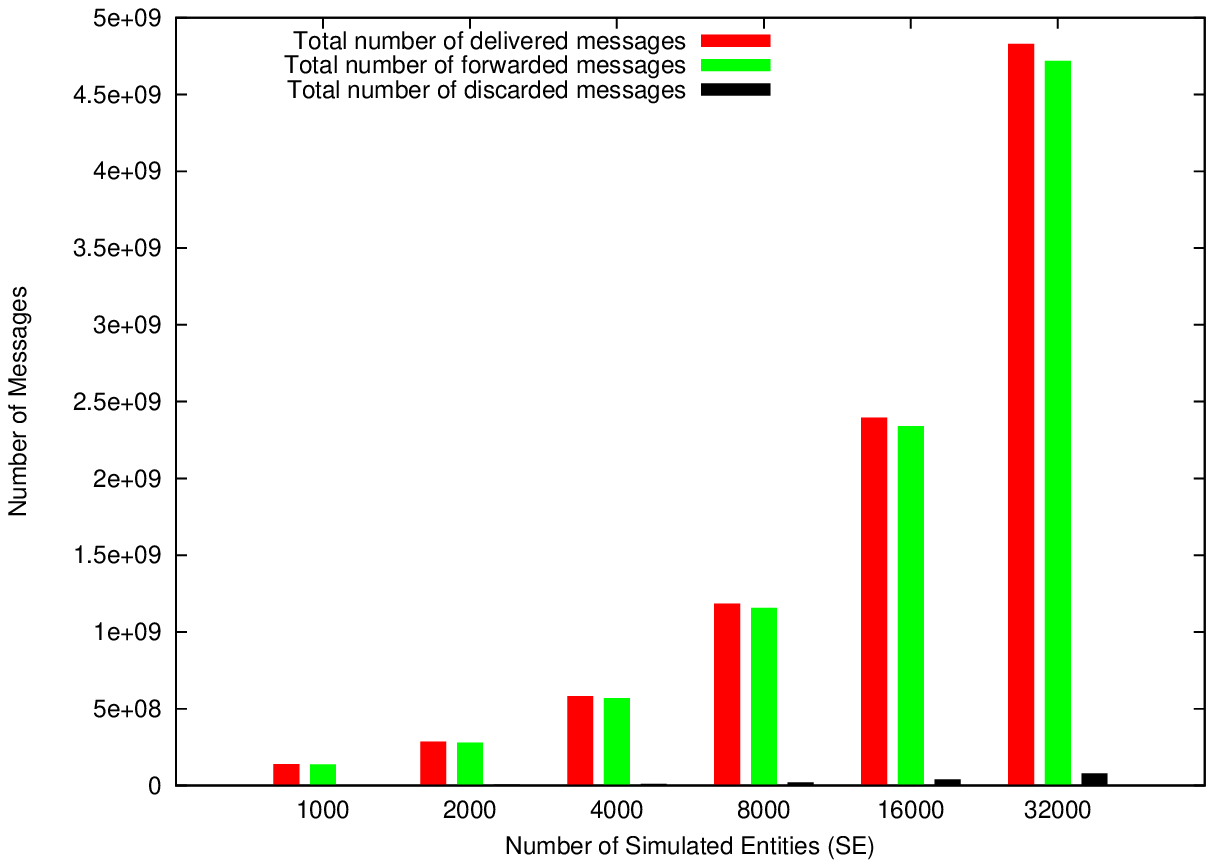}
\caption{Number of delivered messages in a single simulation run (cache ON): increasing number of SEs, total delivered messages, forwarded messages and discarded messages (by the cache)}
\label{fig_messages_cache}
\end{figure}

Figure~\ref{fig_messages_cache} shows that, in this simulation model, the effect of message caching is negligible in terms of reduction of the communication overhead. In fact, only a very small part of the duplicated messages is discarded thanks to caching. This means that, more complex overhead reduction strategies will have to be designed and implemented or different dissemination protocols will have to be employed. Due to the lack of effectiveness of the message caching mechanism, in the rest of this performance evaluation we will always disable the caching mechanism.

Going back to the scalability of the Level 0 simulator, as seen before, the sequential setup shows a limited scalability. An alternative is to implement a parallel simulation~\cite{gda-simpat-2014}. In this case, the simulator uses two or more of the available CPU cores in the execution architecture to process the simulation evolution. The set of SEs was partitioned among the CPU cores and a message passing scheme was used to deliver the interactions among SEs. In this case, the software component executed on each CPU core is called Logical Process (LP).

\begin{figure}[h]
\centering
\includegraphics[width=8.5cm]{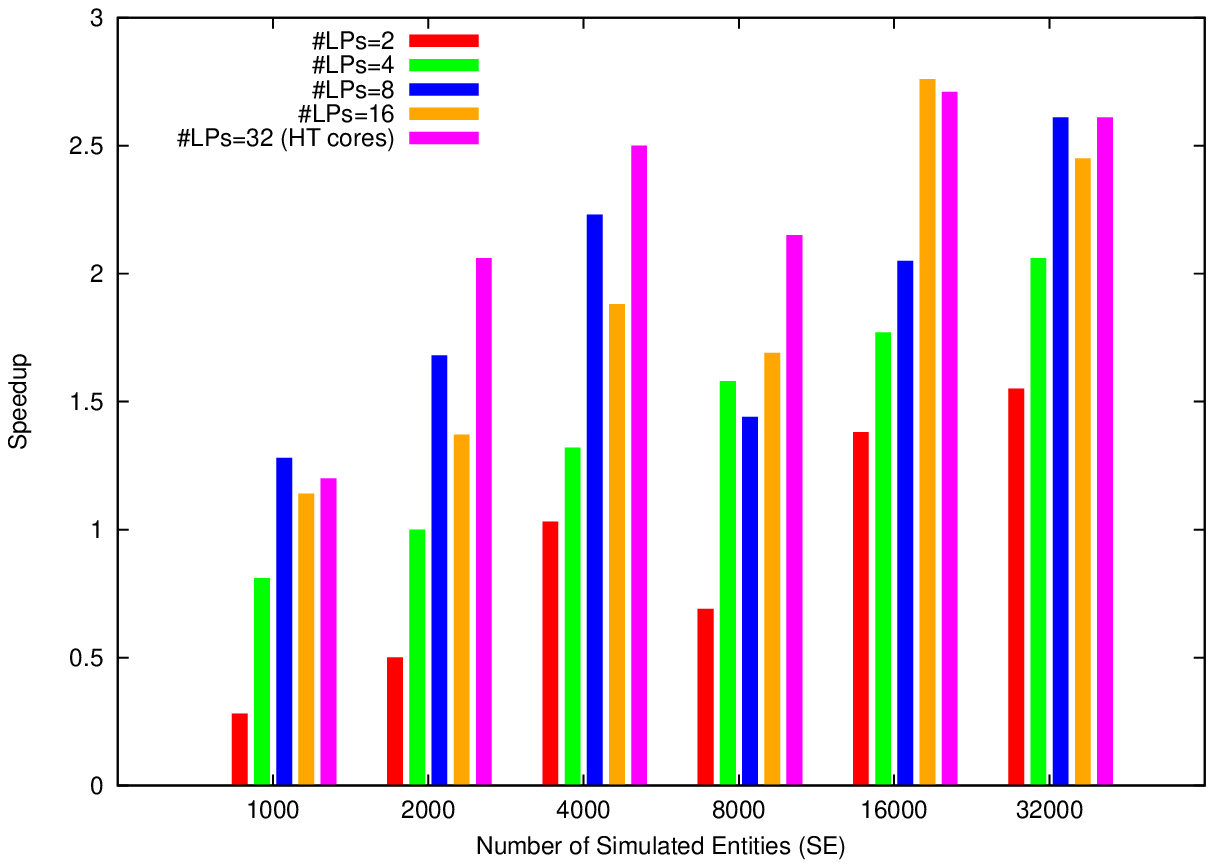}
\caption{Speedup of parallel simulation setup (different number of LPs) with an increasing number of SEs.}
\label{fig_speedupoff}
\end{figure}

Figure~\ref{fig_speedupoff} shows the speedup (i.e.~the ratio between the WCT of the sequential simulation and the WCT of the parallel execution) that can be obtained by GAIA/ART\`IS in different configurations of LPs. In terms of performance, a speedup lower than 1 means that the parallel simulation is slower than the sequential one. On the other hand, a speedup larger than 1 is a gain for the parallel execution. The main advantage of the parallel approach is that it is possible to share the workload among many LPs (and therefore CPU cores). The drawback is the communication costs among LPs.

In the range 1000-8000 SEs (i.e.~moderate loads) there is no speedup for the 2 LPs setup. This means that the communication cost is larger than the benefit of the load sharing of the model computation. The situation is different for larger loads (e.g.~4000 up to 32000), in fact there is always a gain. In general, adding LPs permits a higher speedup. In both scenarios with 16000 and 32000 SEs, the best speedup is obtained when all the physical cores provided by the CPU are used (i.e., 16 cores). In fact, the configuration with 32 LPs is obtained using the CPU logical cores provided by the Hyper-Threading technology. In this case, the Hyper-Threading is unable to improve the performance.

In every parallel setup there is a configuration that has better performance than the sequential one but the speedup is not as good as expected. This is due to the characteristics of the simulated model. In fact, the model is characterized by very little computation performed by each SE and a huge amount of interactions among SEs. With this kind of simulated model, a linear speedup for parallel/distributed execution architectures can not be expected.

For a better assessment of the Level 0 simulator, its memory usage has been measured and analyzed. More in detail, sequential and parallel configurations of the simulator have been evaluated in terms of peak virtual memory usage (VmPeak, i.e., size of the allocated address space) 
and peak resident set size (VmHWM, i.e., used physically resident memory).  
Both measures give insights into the simulator scalability but often with very different outcomes. In fact, in the Linux kernel, a form of lazy memory management is implemented in which the real allocation of memory is deferred until the new memory is actually used~\cite{memory-2}. Tables~\ref{table:VmPeak} and \ref{table:VmHWM} show the VmPeak and VmHWM of the Level 0 simulator with different numbers of LPs and SEs. For each configuration, we show the amount of used memory (KB), as well as the ``overhead'', i.e., the ratio between the memory consumed in the parallel setup (i.e.~$\#LPs>2$) and the memory used in the respective sequential configuration (i.e.~$\#LPs=1$).

As obvious, the amount of used memory increases with the number of SEs and LPs. Up to now, during its development, the $S^3$ simulator has never been optimized in terms of used memory since the main goal is the execution speed. Despite of this, the usage of memory is acceptable even if, for some configurations, the amount of allocated memory is quite large (e.g. Table~\ref{table:VmPeak} with $\#LPs=32$ and $\#SEs=32000$) but the physically memory actually used is much lower. This behavior is mainly due to an implementation that makes a large use of static data structures that are used only in part.

\begin{figure}[h]
\centering
\includegraphics[width=10.5cm]{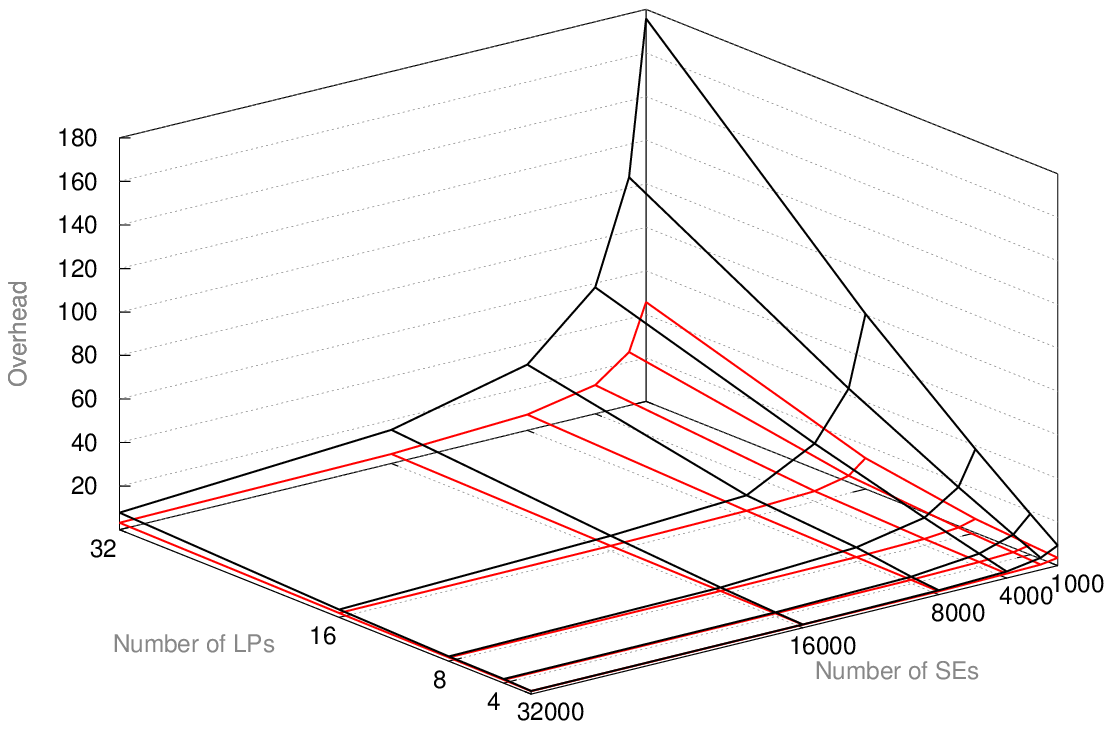}
\caption{Level 0 simulator, ratio between the memory consumed in the parallel setup (i.e.~$\#LPs>2$) and the memory used in the respective sequential configuration (i.e.~$\#LPs=1$). VmHWM in red and VmPeak in black.}
\label{fig_memory_ratio}
\end{figure}

Figure~\ref{fig_memory_ratio} shows the memory overhead (as defined above) for both VmHWM and VmPeak. In general, it is quite limited but it has a sharp increment for configurations with a large number of LPs and few SEs. Anyway, it is worth noting that these configurations are unrealistic, since when the number of SEs is limited then a sequential simulation setup should be preferred. Nevertheless, as future work, the memory usage in the parallel/distributed configuration will be improved.

\begin{table}[]
\scriptsize
\centering
\begin{tabular}{l|c|c|c|c|c|c|c|}
\cline{2-8}
                                             & \multicolumn{7}{c|}{\#SEs}                                                                                              \\ \hline
\multicolumn{1}{|l|}{\multirow{7}{*}{\#LPs}} & \multicolumn{1}{l|}{} & \textbf{1000} & \textbf{2000} & \textbf{4000} & \textbf{8000} & \textbf{16000} & \textbf{32000} \\ \cline{2-8} 
\multicolumn{1}{|l|}{}                       & \textbf{1}            & 40192 ($1$)             & 67440 ($1$)             & 121876 ($1$)             & 234640 ($1$)             & 459720 ($1$)              & 913588 ($1$)              \\ \cline{2-8} 
\multicolumn{1}{|l|}{}                       & \textbf{2}            & 371744 ($9.25$)             & 372016 ($5.52$)             & 372296 ($3.05$)             & 465792 ($1.99$)             & 694728 ($1.51$)              & 1286312 ($1.41$)              \\ \cline{2-8} 
\multicolumn{1}{|l|}{}                       & \textbf{4}            & 753104 ($18.74$)             & 753088 ($11.17$)             & 753616 ($6.18$)             & 754960 ($3.22$)             & 950320 ($2.07$)              & 1678512 ($1.84$)              \\ \cline{2-8} 
\multicolumn{1}{|l|}{}                       & \textbf{8}            & 1543712 ($38.41$)             & 1543360 ($22.88$)             & 1544416 ($12.67$)             & 1547328 ($6.59$)             & 1551456 ($3.37$)              & 1993312 ($2.18$)              \\ \cline{2-8} 
\multicolumn{1}{|l|}{}                       & \textbf{16}           & 3233728 ($80.46$)             & 3238464 ($48.02$)             & 3240512 ($26.59$)             & 2436992 ($10.39$)             & 3252672 ($7.08$)              & 3267200 ($3.58$)              \\ \cline{2-8} 
\multicolumn{1}{|l|}{}                       & \textbf{32}           & 7067520 ($175.84$)             & 7072128 ($104.87$)             & 7081216 ($58.10$)             & 7077248 ($30.16$)             & 7091968 ($15.43$)              & 7134592 ($7.81$)              \\ \hline
\end{tabular}
\caption{Level 0 simulator, peak virtual memory usage (VmPeak) in KB. The ratio between the memory consumed in a given parallel setup and the memory used in the sequential configuration is reported.}
\label{table:VmPeak}
\end{table}

\begin{table}[]
\scriptsize
\centering
\begin{tabular}{l|c|c|c|c|c|c|c|}
\cline{2-8}
                                             & \multicolumn{7}{c|}{\#SEs}                                                                                              \\ \hline
\multicolumn{1}{|l|}{\multirow{7}{*}{\#LPs}} & \multicolumn{1}{l|}{} & \textbf{1000} & \textbf{2000} & \textbf{4000} & \textbf{8000} & \textbf{16000} & \textbf{32000} \\ \cline{2-8} 
\multicolumn{1}{|l|}{}                       & \textbf{1}            & 29564 ($1$)             & 56840 ($1$)             & 111416 ($1$)             & 223908 ($1$)             & 449068 ($1$)              & 903156 ($1$)              \\ \cline{2-8} 
\multicolumn{1}{|l|}{}                       & \textbf{2}            & 105416 ($3.57$)             & 138784 ($2.44$)             & 230624 ($2.07$)             & 351600 ($1.57$)             & 635880 ($1.42$)              & 1208936 ($1.34$)              \\ \cline{2-8} 
\multicolumn{1}{|l|}{}                       & \textbf{4}            & 132880 ($4.49$)             & 171728 ($3.02$)             & 263488 ($2.36$)             & 396416 ($1.77$)             & 714720 ($1.59$)              & 1351072 ($1.59$)              \\ \cline{2-8} 
\multicolumn{1}{|l|}{}                       & \textbf{8}            & 187712 ($6.35$)             & 242176 ($4.26$)             & 327584 ($2.94$)             & 476480 ($2.13$)             & 815104 ($1.82$)              & 1521056 ($1.68$)              \\ \cline{2-8} 
\multicolumn{1}{|l|}{}                       & \textbf{16}           & 418176 ($14.14$)             & 456704 ($8.03$)             & 521664 ($4.68$)             & 734144 ($3.28$)             & 1088576 ($2.42$)              & 1820736 ($2.02$)              \\ \cline{2-8} 
\multicolumn{1}{|l|}{}                       & \textbf{32}           & 1349120 ($45.63$)             & 1395072 ($24.54$)             & 1466112 ($13.16$)             & 1621760 ($7.24$)             & 1955840 ($4.36$)              & 2923136 ($3.24$)              \\ \hline
\end{tabular}
\caption{Level 0 simulator, peak resident set size (VmHWM) in KB. The ratio between the memory consumed in a given parallel setup and the memory used in the sequential configuration is reported.}
\label{table:VmHWM}
\end{table}

Finally, the adaptive partitioning supported by GAIA/ART\`IS has been activated. The goal of the mechanism is to cluster the SEs with the aim to reduce the amount of communication between the LPs. More in detail, SEs that are near in the simulated area are clustered in the same LP. This approach, that is based on the analysis of the interaction pattern of each SE, is often able to reduce the communication cost and hence to speedup the simulation runs. Furthermore, it can be also extended to deal with computational imbalances in the execution architecture (e.g.~background load) or in the simulation model (e.g.~hotspots).

\begin{figure}[h]
\centering
\includegraphics[width=8.5cm]{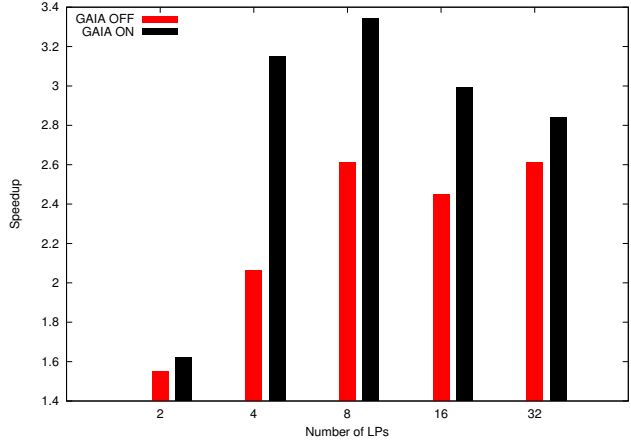}
\caption{Speedup of adaptive parallel simulation setup (different number of LPs, 32000 SEs).}
\label{fig_speedupon}
\end{figure}

Figure~\ref{fig_speedupon} studies the effect of the adaptive mechanism on 32000 SEs and in presence of an increasing number of LPs. The adaptive mechanism (in red) always gives a gain with respect to static partitioning (in black). The gain is limited for 2 LPs but increases in the larger setups. The best results (that is 3.34 vs. 2.61) is obtained with 8 LPs while with 16 and 32 LPs the gain is lower. This behavior can be explained as follows: up to 8 LPs the load parallelization gives a speedup. With a higher number of LPs, the load parallelization is exceeded by the extra cost for communication. This means that when 8 LPs are used, the computational load is properly partitioned, the adaptive mechanism is able to reduce the communication cost and hence it obtains a significant performance gain.

%%%%%%%%%%%%%%%%%%%%%%%%%%%%%%%%%%%%%%%%%%%%%%%%%%%
\subsection{Level 1: OMNeT++ simulator}
%%%%%%%%%%%%%%%%%%%%%%%%%%%%%%%%%%%%%%%%%%%%%%%%%%%

In this section, the performance of the finer grained simulator described in Section~\ref{sec:omnet} is evaluated to assess its scalability. As mentioned above, this Level 1 model has been implemented using OMNeT++ and, in our opinion, it is a good example of domain specific simulator. In this case, our goal is twofold. Firstly, we provide evidence that widely used domain specific simulators are unable to scale to the degree required by IoT deployments. Secondly, we aim to evaluate the simulator that has been integrated in our multi-level simulator before studying the performance of the integrated tool.

\begin {table}[h]
\begin{center}
    \begin{tabular}{| l | c | c | c |}
    \hline
    $\#SEs$ & VmPeak (KB) & VmHWM (KB) & WCT (sec) \\ \hline
    1     &   147976    &   43916	&  $14.878$ \\ \hline
    10    &   151476    &   47352	&  $114.842$ \\ \hline
    100   &   198380    &   94204	&  $2012.684$ \\ \hline
    1000  &   2069432   &   1958456	&  $34103.106$ \\ \hline
    10000 &  $>20694320$ &  $>19584560$	&   $>3$ days \\ \hline
    \hline
    \end{tabular}
	\caption{Performance assessment of the Level 1: OMNeT++ simulator. VmPeak is the peak virtual memory usage and VmHWM is the peak resident set size.}
	\label{table:omnet}
\end{center}
\end {table}

Table~\ref{table:omnet} shows that the amount of memory used by OMNeT++ significantly increases with the number of SEs. Furthermore, the WCT required by the simulator to complete each run sharply increases. In all tested cases, the run length of the Level 1 OMNeT++ model is set equal to the timestep of the Level 0 simulator (as described in Section~\ref{sec:multilevel}). This confirms that, in this specific simulated model, the OMNeT++ scalability is adequate for the simulation of small size systems but it does not fit with the requirements of large scale IoT deployments.

%%%%%%%%%%%%%%%%%%%%%%%%%%%%%%%%%%%%%%%%%%%%%%%%%%%
\subsection{Multi-level simulator}
%%%%%%%%%%%%%%%%%%%%%%%%%%%%%%%%%%%%%%%%%%%%%%%%%%%

In this section, the performance of the prototypal multi-level simulator are assessed. First of all, it is considered a sequential Level 0 simulation (with 1000 SEs and $\#LPs=1$) in which, a given number of sequential OMNeT++ instances are spawned. As described in Section~\ref{sec:multilevel}, at given points in time, the Level 0 simulator triggers the execution of a more detailed simulation for a specific zone of the simulated area. In this performance evaluation, we consider a setup in which a single SE is transfered from the Level 0 simulator and it is managed by the Level 1 for the length of a single Level 0 timestep. The results of this experiment are reported in Table~\ref{table:multilevel1000} and they show that the total WCT of the multi-level simulator increases linearly with the number of Level 1 simulator spawns. As expected, the cost added by spawning an OMNeT++ is in line with what reported in Section~\ref{table:omnet} and the interoperability overhead introduced by the coordination between the two simulators (i.e.~Level 0 and Level 1) is negligible.

The amount of memory used by the multi-level simulator is given by the memory used by the Level 0 simulator and the allocations caused by the Level 1 instances. Even if the impact of a single Level 1 instance is acceptable, it is worth noting that, in some cases it may be necessary to run multiple concurrent Level 1 instances. This means that, the peak used memory of the multi-level simulator can be very high and that, in these cases, the distributed simulator configuration (based on multiple interconnected hosts) is to be preferred to the sequential or parallel setups.

\begin {table}[h]
\begin{center}
    \begin{tabular}{| c | c | c | c |}
    \hline
    \# spawns & WCT (multi-level) & WCT (Level 0) & WCT per spawn \\ \hline
    1	&	52.31	&	37.69	&	14.62	\\ \hline
    2	&	66.825	&	37.69	&	14.56	\\ \hline
    3	&	79.68	&	37.69	&	14.00	\\ \hline
    4	&	93.16	&	37.69	&	13.86	\\ \hline
    5	&	108.12	&	37.69	&	14.08	\\ \hline
    6	&	122.35	&	37.69	&	14.11	\\ \hline
    7	&	135.81	&	37.69	&	14.01	\\ \hline
    8	&	150.94	&	37.69	&	14.15	\\ \hline
    \hline
    \end{tabular}
	\caption{Performance assessment of the multi-level simulator, sequential ($\#LPs=1$) configuration with 1000 SEs.}
	\label{table:multilevel1000}
\end{center}
\end {table}

Another aspect worth of consideration is the number of SEs managed by the Level 0 simulator. We have measured the WCT of the multi-level simulator, the Level 0 simulator (Table~\ref{table:multilevelSE} columns 1 and 2) and computed the difference (that is $\Delta$WCT). $\Delta$WCT represents the overhead (in terms of WCT) that has been introduced by the spawning of Level 1 simulator instances. As shown in the table, increasing the number of SEs managed by the upper level simulator has the effect to reduce the $\Delta$WCT. This is a counterintuitive behavior but it is compliant with the design of the multi-level simulator.

In fact, $\Delta$WCT is basically the time spent by Level 0 waiting for results from the finer grained level. This means that the Level 0 simulator has computed its coarse simulation step needed to advance its simulation, apart from updates coming from the finer level (Level 1). In this case, the bottleneck is the lower level simulator. In general, at every Level 0 timestep, the higher the amount of Level 0 simulated entities the higher the workload at the higher level, thus the lower the time spent waiting for Level 1 advancements.

\begin{figure}[ht]
\centering
\subfloat[Small \#SEs, larger idle time.]{\includegraphics[width=11cm]{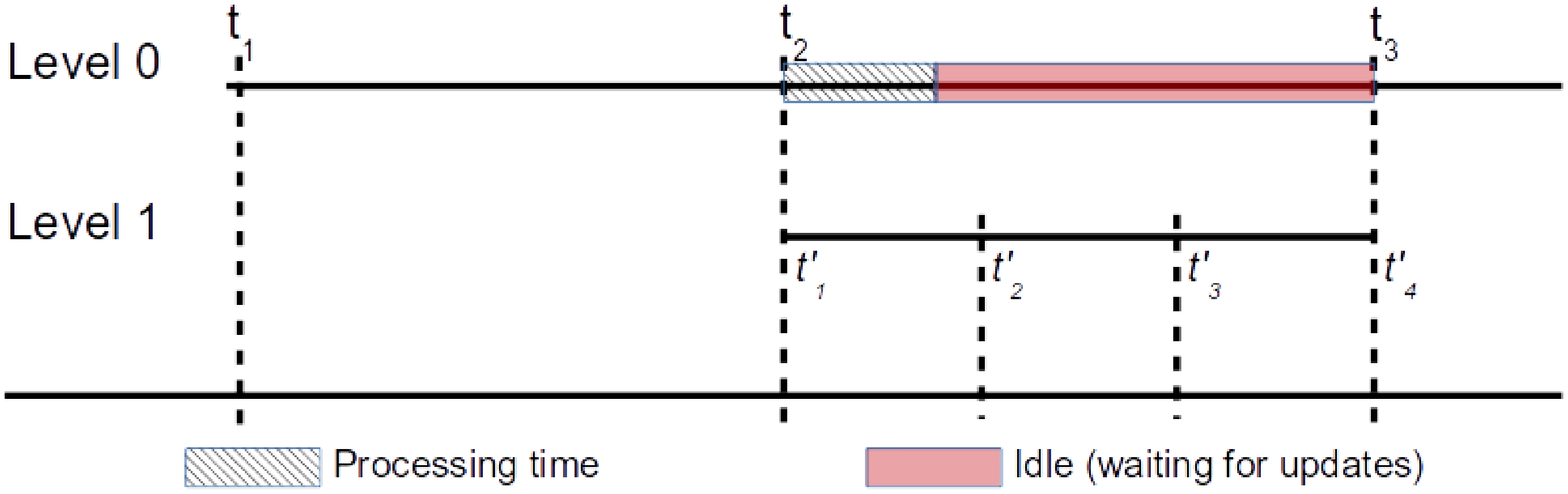}\label{fig:multilevelSE_load_2}}\\
\subfloat[Large \#SEs, smaller idle time.]{\includegraphics[width=11cm]{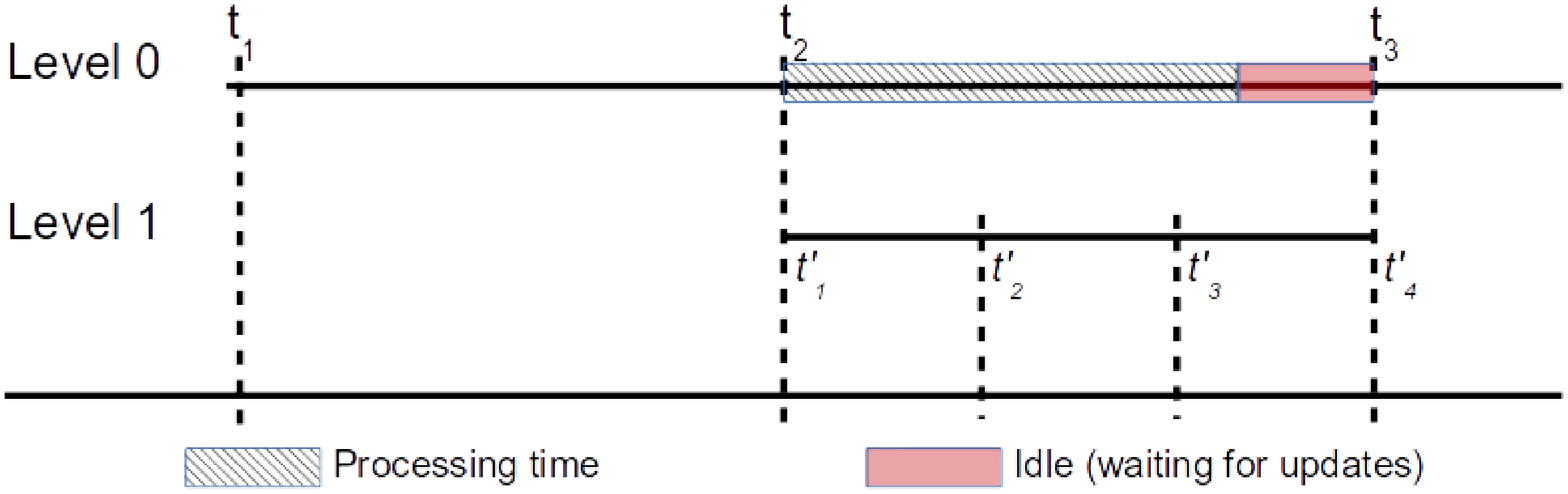}\label{fig:multilevelSE_load_1}}
\caption{Multi-level simulator: Level 0 allocates a number of SEs, it must block waiting for completion of the Level 1 simulator.}
\end{figure}

In other words, with a low number of SEs, the Level 0 simulator is so fast in processing its coarse size timesteps that when a Level 1 instance is spawned then it has to block up to when the instance ends its computation and reports the results (see Figure~\ref{fig:multilevelSE_load_2}). Conversely, this does not happen when the number of SEs is large. In fact, in this case, for the most part of the Level 0 timestep both the Level 0 and Level 1 simulator can run in parallel (see Figure~\ref{fig:multilevelSE_load_1}). This concurrent execution of the Level 0 coarse grained timestep and the fine grained Level 1 simulator has the effect to reduce the overhead introduced by the spawn of Level 1 instances.

\begin {table}[h]
\begin{center}
    \begin{tabular}{| c | c | c | c |}
    \hline
    \#SEs & WCT (multi-level) & WCT (Level 0) & $\Delta$WCT \\ \hline
    1000	&	52.31	&	37.69	&	14.62	\\ \hline
    4000	&	278.68	&	263.87	&	14.81	\\ \hline  
    8000	&	804.20	&	793.20	&	11.00	\\ \hline
    \hline
    \end{tabular}
	\caption{Performance assessment of the multi-level simulator, sequential configuration ($\#LPs=1$) with an increasing number of SEs.}
	\label{table:multilevelSE}
\end{center}
\end {table}

In the parallel/distributed configuration ($\#LPs>1$), the performance of the multi-level simulator is dominated by the behavior of the Level 0 simulator. In other words, in most configurations the spawning of Level 1 instances results as a slow down of the Level 0 simulator. As shown in Figures~\ref{fig:multilevelSE_load_1} and~\ref{fig:multilevelSE_load_2}, when the Level 0 computation is larger than in the Level 1 there is some balancing. More often, the overhead introduced by Level 1 results in a slow down of the LP that has spawned the Level 1 instance. In other words, this specific LP becomes the bottleneck of the whole parallel/distributed simulation and some computational and communication imbalances are introduced in the execution architecture.

\begin{figure}[h]
\centering
\includegraphics[width=8.5cm]{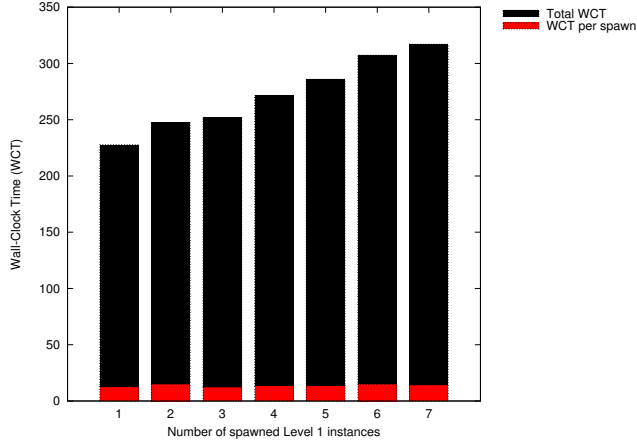}
\caption{WCT of the parallel multi-level simulator with $\#LPs=4$, 4000 SEs, increasing number of spawned Level 1 instances (in sequence by a single LP).}
\label{fig:pads-histo-tot_activations}
\end{figure}

To better assess the performance of the multi-level simulator in a parallel setup, we have firstly studied the behavior of the simulator when 4 LPs are used and in which a single specific LP spawns a series of Level 1 instances. As before, the finer grained instances are executed by the LP in sequence during the simulation run.  Figure~\ref{fig:pads-histo-tot_activations} shows the total WCT and how much time the spawning of each Level 1 instance has added to the total time of execution (i.e.~WCT per spawn). As expected, the total WCT increases linearly with the number of spawns and the cost of each spawn is almost constant.

\begin{figure}[h]
\centering
\includegraphics[width=8.5cm]{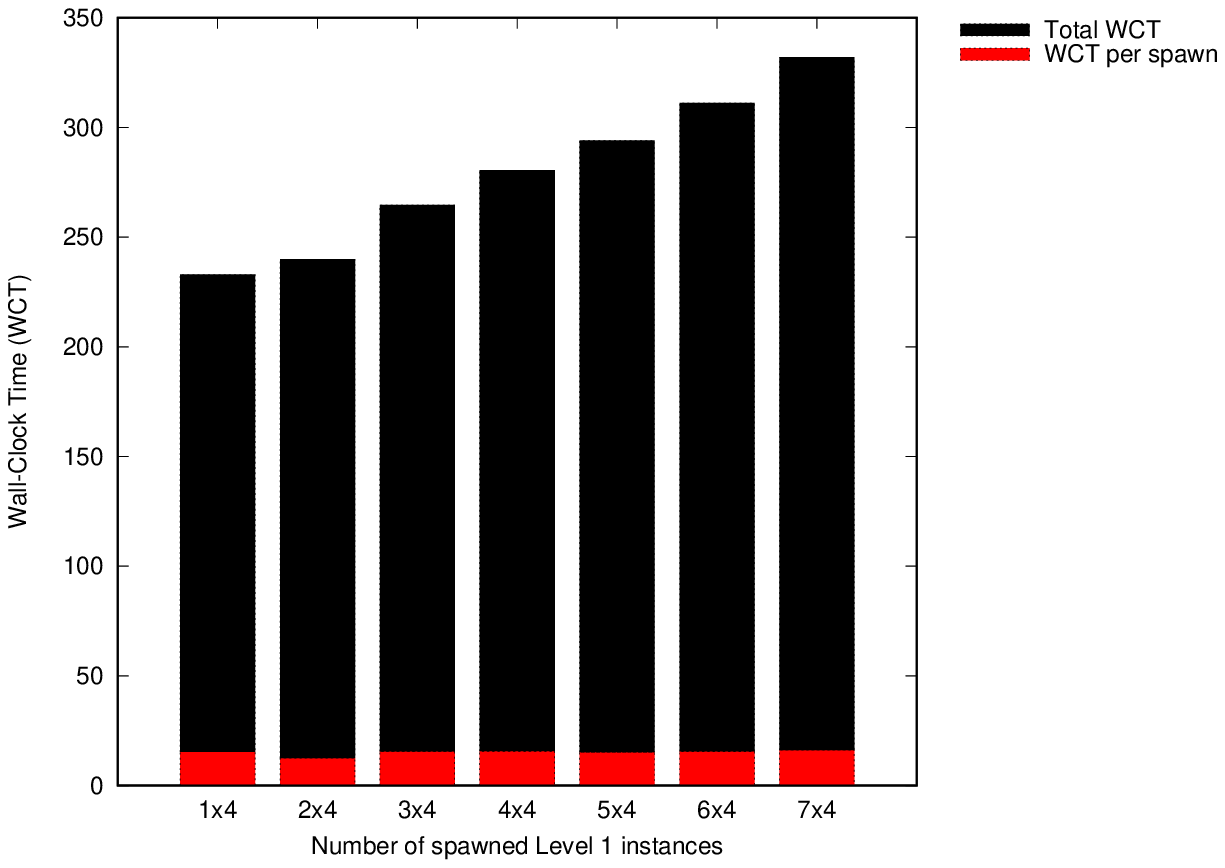}
\caption{WCT of the parallel multi-level simulator with $\#LPs=4$, 4000 SEs, increasing number of concurrent spawned Level 1 instances. For example, 2x4 means that all the LPs for 2 times during the simulation run trigger 4 concurrent Level 1 instances.}
\label{fig:pads-histo-multiple_activations}
\end{figure}

The same setup has been used to investigate what happens if all the LPs in the parallel execution architecture spawn a Level 1 instance at given points in the simulated time. This means that the total number of triggered instances in the simulation run is much higher and that multiple concurrent Level 1 instances are run. Figure~\ref{fig:pads-histo-multiple_activations} shows the WCT of the multi-level simulator with an increasing number of concurrent instances. The goal of this test is to demonstrate that while increasing the number of activations of lower level instances is costly (the WCT increases as the number of times that the instances are triggered), there is no extra cost at running concurrent instances in different CPU cores. In fact, if idle CPU cores are available then all the concurrent instances can be run in parallel. In the case of this experiment, this can be verified comparing the WCT per spawn reported in Figure~\ref{fig:pads-histo-multiple_activations} and in Figure~\ref{fig:pads-histo-tot_activations}.

\begin{figure}[h]
\centering
\includegraphics[width=8.5cm]{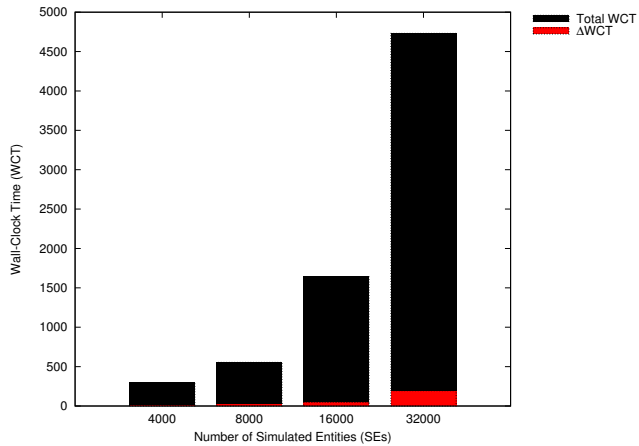}
\caption{WCT of the parallel multi-level simulator with $\#LPs=4$, single Level 1 spawn, increasing number of SEs.}
\label{fig:pads-histo-increasing_SEs}
\end{figure}

Going back to the configuration in which a single Level 1 instance is spawned at a given time by a specific LP, an interesting and counterintuitive result is obtained when the number of SEs is varied. Figure~\ref{fig:pads-histo-increasing_SEs} reports the total WCT of the multi-level simulation and the $\Delta$WCT (as defined before). It is in line with expectations that the total WCT increases with the number of SEs but it is counterintuitive that also the $\Delta$WCT is affected by the number of SEs. As mentioned above, in this performance evaluation, a single SE is transfered from the Level 0 simulator and it is managed by the Level 1 for the length of a single Level 0 timestep. In other words, the Level 1 spawn is independent from the number of SEs managed in Level 0. Since $\Delta$WCT represents the overhead (in terms of WCT) that has been introduced by the spawning of Level 1 simulator instances, then it is counterintuitive that this overhead changes with respect to an instance that is constant in terms of length and complexity. The reason of this behavior is simple. In fact, the spawning of a Level 1 instance results in a temporary loss of coordination in the parallel/distributed execution architecture. This mean that, due to the synchronization mechanism, all the LPs except the one that has spawned the instance will be blocked waiting for its completion of the current Level 0 timestep. Since the LP that has spawned the instance will be late, then some operations needed for managing the end of timestep have to be completed while all the other parts of the parallel/distributed execution architecture are waiting. These end of step operations depend on the number of SEs managed by the LP and therefore also the $\Delta$WCT is influenced by the number of SEs.

%%%%%%%%%%%%%%%%%%%%%%%%%%%%%%%%%%%%%%%%%%%%%%%%%%%
\subsection{Discussion}
%%%%%%%%%%%%%%%%%%%%%%%%%%%%%%%%%%%%%%%%%%%%%%%%%%%

Results obtained from our simulations confirm the viability of the proposal. The use of multi-level simulation allows scaling up to higher numbers of SEs, with respect to the use of a single fine-grained simulator that is able to capture the complexity and all the technical details of the interactions among SEs. 
This approach allows mimicking all these details only when needed. Hence, during the rest of the simulation the multi-level simulator behaves as a coarse-grained simulator.

The use of an adaptive and parallel approach that distributes the execution of SEs in multiple LPs (and CPU cores) shows benefits, provided that interacting SEs are clustered into the same LP. Indeed, the best case scenario for this approach is when a simple partitioning of SEs is possible. For example, imagine having separated simulated areas (because of distance, obstacles, walls) that ease the partitioning and the clustering of SEs. In this case, a migration of a SE from a cluster to another is needed only when the SE moves from a simulated area to another.

Conversely, the worst case scenario is in presence of an hotspot, where the majority of SEs stays in the same (restricted) simulated area. In this case, finding a good partitioning might be a difficult task. Moreover, in presence of large size hotspots, it would be complex to determine at which level of detail each SE needs to be simulated and to guarantee the necessary isolation of the Level 0 simulator from the Level 1.

Another issue that might complicate a multi-level simulation such that proposed in this work is when the higher level simulation (Level 0 in the scenario presented above) exploits very short timesteps (with respect to the actions being simulated), making difficult splitting them into finer grained timesteps at lower levels.
Finally, it is clear that the higher the number of finer grained level instances to be triggered during the simulation, the higher the complexity of the simulation and thus the higher the time to accomplish it.

\section{Related Work}
\label{sec:related}

The use of IoT to build ``smart'' services in territories raises several issues related to the design, implementation and deployment of these services in real environments. 
In fact, even a small size smart territory will be composed by thousands of interconnected devices. Many of them will be mobile and each with very specific behavior and technical characteristics~\cite{smartshires}. 
Simulation provides means to better approach this issue.
If a proactive approach is needed (e.g.~simulation in the loop), in order to perform ``what-if analysis'' during the management of the deployed architecture, then the simulator should be able to run in (almost) real-time, at least with average size model instances. 

\subsection{Simulation of the Internet of Things}

IoT setups require simulating large scale testbeds producing a huge amount of data, at a constant rate. This requires the use of scalable simulation tools, able to support a massive amount of nodes in the scenario and fine level of detail in the nodes interactions.
This complicates the simulation.

Clearly enough, the need for scalability and high level of detail make existing network simulators, such as OMNeT++, ns-2 or ns-3, quite often inadequate for the purpose, when used alone~\cite{6069710}.
On the other hand, agent-based simulation is a perfect tool to create models that mimic wide area (e.g.~urban) systems in general, that can be applied at different time scales, such as short-term modeling, e.g. diurnal patterns in cities, and long-term models for exploring change through strategic planning~\cite{Karnouskos}. Tools such as MASON~\cite{Luke:2005} and SUMO~\cite{SUMO2012} allow simulating moving entities (e.g.~mobile users of vehicles) that can interact with static ones. These tools have been successfully exploited to study intelligent traffic control systems~\cite{bauza,kerekes,Wegener:2008,e16052384}, mobile applications that resort to crowdsensed data 
and so on. The main problem of these approaches is that, due to their nature, they do not allow considering high levels of details.

Probably, the most common approach to simulate IoT is that of resorting to discrete event simulation approaches. However, different types of simulation have been considered to simulate IoT. 
For instance, 
MAMMotH is a software architecture based on emulation~\cite{6664581}.
Monte Carlo methods are employed in~\cite{6418824}.

Model-driven simulation (based on the standard language SDL) is used to describe an IoT scenario~\cite{Brumbulli2016}. Then, an automatic code generation transforms the description into an executable simulation model for the ns-3 network simulator.

Brambilla et al. propose to integrate the (monolithic, Java-based) DEUS general-purpose discrete event simulation with the domain specific simulators Cooja and ns-3 for implementing large-scale IoT simulations in urban environments~\cite{Brambilla:2014:SPL:2694768.2694780}. In this case, the performance evaluation is based on 6 scenarios with up to $200000$ sensors, $400$ hubs and $25000$ vehicles. The execution time with respect to the number of events shows a quite good scalability. 

DPWSim is a simulation toolkit that supports the modeling of the OASIS standard ``Devices Profile for Web Services'' (DPWS)~\cite{6803226}. Its main goal of is to provide a cross-platform and easy-to-use assessment of DPWS devices and protocols. In other words, it is not designed for very large-scale setups.

The use of cloud computing can provide better performance. SimIoT exploits cloud environments for back-end operations~\cite{6844677}. The use case proposed in the paper is a health monitoring system for emergency situations in which short range and wireless communication devices are used to monitor the health of patients. The preliminary performance evaluation is based on 160 identical jobs submitted by 16 IoT devices.

Finally, an interesting approach is described in~\cite{paper+kirsche-13:iot-simulation}. The author proposes a hybrid simulation environment in which the Cooja-based simulations (i.e.~system level) are integrated with a domain specific network simulator (i.e.~OMNeT++).
In some sense, this approach resembles the idea of exploiting a multi-level simulation, since it tries to keep the best of the considered simulation approaches.

To sum up, we claim that a multi-level simulation is needed in order to simulate reasonable IoT scenarios. In fact, running the whole model at the highest level of detail is unfeasible. A better approach is to bind different simulators together, each one running at its appropriate level of detail and with specific characteristics of the domain to be simulated (e.g.~mobility models, wireless/wired communications and so on).

\section{Conclusions}
\label{sec:conclusions}

In this paper, we have presented a principal instance of multi-level simulation of the Internet of Things (IoT).
The IoT is going to be composed of billions of devices, generating a massive amount of data at a constant pace. This makes such a case study particularly complex in terms of design and modeling.
Simulation must be carefully handled in order to create reasonably accurate models that can scale in terms of modeled entities and granularity of events. The paradigm of multi-level simulation enables such scalable and fast simulations, since fine level of details are employed when needed, only. In other contexts, more coarse grained simulation scenarios are employed. 

Our specific solution uses a two level simulation. An adaptive, parallel/distributed agent-based simulation technique is employed to model the coarse level (Level 0), while an OMNeT++ simulation implements the finer level (Level 1). During the simulation, the coarse Level 0 is in charge of coordinating the different instances at Level 1 that may have been triggered in different portions of the simulated world. In fact, Level 0 can decide when starting a Level 1 simulation, and at every time-step it decides if such Level 1 simulation should continue or if it must stop.

Our multi-level simulator has been specifically utilized to study novel solutions fostering the creation of smart services for countrysides. The underlying idea is that decentralized territories need smart services that must be cheap, adaptive, self-configuring and robust. As an application scenario, we considered a smart market, with customers that can subscribe their interest to specific products that may become available in some location of their neighborhood. Upon availability, the customer can travel to the market. Once there, additional services can be provided to the users, such as detailed information, advertisements, guidance through the market towards the location of the producer.
The coarse grained (Level 0) simulator models the publish/subscribe scheme and the wide area movements, while the finer grained (Level 1) simulator implements all the interactions and wireless communications
occurring within the smart market.
Based on this use case, it is confirmed that the ad-hoc wireless networking technologies do represent a principle tool to deploy smart services over decentralized countrysides.
Moreover, we provided a performance evaluation that confirms the viability of utilizing multi-level simulation for simulating large scale IoT environments.

%%%%%%%%%%%%%%%%%%%%%%%%%%%%%%%%%%%%%%%%%%%%%%%%%%%%%%%%%%%%%%%%%%%%%%%%%%%%%%%%%%%%%
\bibliographystyle{elsarticle-num}
\bibliography{paper}
%%%%%%%%%%%%%%%%%%%%%%%%%%%%%%%%%%%%%%%%%%%%%%%%%%%%%%%%%%%%%%%%%%%%%%%%%%%%%%%%%%%%%

\end{document}